\documentclass[a4paper,12pt,nohyper]{JHEP3}
\usepackage{amsmath,latexsym,amssymb}
\usepackage{graphicx}
\usepackage[latin1]{inputenc}
\usepackage{subfigure}
\usepackage{nicefrac}
\usepackage{braket}
\usepackage{pifont}

\RequirePackage{graphicx}

\usepackage{framed}

\def\crbig{\\\noalign{\vspace{1mm}}}

\def\r{\rho}

\def\G{\Gamma}

\def\D{\Delta}
\def\e{\epsilon}

\def\th{\theta}

\def\m{\mu}
\def\n{\nu}
\def\om{\omega}

\def\no{\noindent \vspace{-0.35 cm}\newline}

\def\qq{\qquad}

\def\IR{\relax{\rm I\kern-.18em R}}

\def \ha {{1\over 2}}

\def \ov {\over}

\def\IR{\relax{\rm I\kern-.18em R}}
\def\inv{^{\raise.15ex\hbox{${\scriptscriptstyle -}$}\kern-.05em 1}}

\def\be{\begin{equation}}
\def\ee{\end{equation}}
\def\ba{\begin{eqnarray}}
\def\ea{\end{eqnarray}}

\def\del{\partial}

\newcommand{\eqn}[1]{(\ref{#1})}

\title{\boldmath\Large
Supersymmetric deformations of
F1-NS5-branes and their exact CFT description
\unboldmath}

\author{
P.M.~Petropoulos${}^{1}$\footnote{marios@cpht.polytechnique.fr},
N.~Prezas${}^{2}$\footnote{prezas@itp.unibe.ch} and
K.~Sfetsos${}^{3}$\footnote{sfetsos@upatras.gr}
\\

  \begin{itemize}

  \item     Centre de Physique Th{\'e}orique, Ecole Polytechnique, CNRS--UMR 7644,\\
91128 Palaiseau Cedex, France

\item Institute for Theoretical Physics, University of Bern,\\
3012 Bern, Switzerland

 \item Department of Engineering Sciences, University of Patras,\\
26110 Patras, Greece

        \end{itemize}

}

\abstract{ We consider certain classes of operators in the exact
conformal field theory $SL(2,\mathbb{R}) \times SU(2) \times U(1)^4$
describing strings in an $\mathrm{AdS}_3 \times S^3 \times \mathbb{T}^4$
geometry supported by Neveu--Schwarz 3-form fluxes. This background arises
in the near-horizon limit of a system of NS5-branes wrapped on a
4-torus and F1-branes smeared on the 4-torus when both types of
branes are located at the same point in their common transverse space.
We find a class of operators that lead to spacetime supersymmetric
deformations. It is remarkable that most of these operators are not
chiral primary with respect to the ${\cal N}=2$ superconformal
algebra on the worldsheet. A subset of these worldsheet conformal field theory
deformations admits
an interpretation either as a geometric deformation of the brane
system or as a deformation of the distribution of
the F1-branes,  viewed as smooth instantons,  inside the wrapped
NS5-brane worldvolume. The 2-dimensional conformal field theory, however,  seems to
lack operators corresponding to arbitrary NS5-brane deformations, in
contrast to pure NS5-brane systems where all geometric deformations
can be accounted for by chiral primary operators.

  }

\preprint{CPHT-RR020.0409}

\begin{document}

\section{Introduction}

Objects charged under the NSNS antisymmetric tensor field of string theory,
i.e.~the electrically charged F1-branes and the magnetically charged NS5-branes
as well as their bound states,
are of particular importance since the corresponding string theory backgrounds may in principle
admit
an exact conformal field theory (CFT) description. In such cases
the physics of these objects is amenable to the powerful methods of
CFT.

\no The most well-known example is that of a configuration of parallel and coincident
NS5-branes. Their backreaction leads to a characteristic throat-like geometry
whose near-horizon limit comprises of a linear dilaton along with a
3-sphere, both of which can be described in terms of exact CFTs
\cite{Callan:1991dj}.  Another example is provided by a circular distribution
of NS5-branes. This system, after an appropriate T-duality, admits a CFT
description in terms of the cosets ${SU(2)}/{U(1)} \times
{SL(2,\mathbb{R})}/{U(1)}$
\cite{Sfetsos:1998xd} and can be thought of as a
deformation of the first configuration that resolves the strong coupling
singularity associated with the linear dilaton \cite{Giveon:1999px, Aharony:2003vk}.

\no An interesting feature of these systems is that the little string theories (LSTs)
that reside on the worldvolume of the NS5-branes \cite{Seiberg:1997zk} can be described holographically
in terms of the associated CFTs \cite{Aharony:1998ub}. A fundamental aspect
of these holographic dualities is the existence of a dictionary between
deformations of the branes described via perturbations of the
original supergravity solution, parametrized by vacuum expectation values
of scalar fields on the branes, and exactly marginal
 deformations of the underlying CFT. Such a dictionary was discussed
 in detail in  \cite{Aharony:2003vk} and was tested successfully  in \cite{Fotopoulos:2007rm}, by matching
 directly the supergravity deformations realized in the  $\sigma$  model description of the theory
 to CFT operators.

 \no The latter analysis
 was motivated by earlier work  \cite{Marios Petropoulos:2005wu}
where it was explicitly shown that the continuous deformation of the circular
NS5-brane distribution into an elliptic one was driven by a
marginal perturbation of the ${SU(2)}/{U(1)} \times
{SL(2,\mathbb{R})}/{U(1)}$ worldsheet $\sigma$ model. The deformation of the circle into an ellipsis  is one particular mode among an infinitude consisting of battered
circles with $n\in \mathbb{N}$ bumps distributed with
$\mathbb{Z}_n$ symmetry around the original circle. These types of deformations as well as
their corresponding CFT operators based on parafermions provided actually the testing
ground for  \cite{Fotopoulos:2007rm}.

\no One interesting aspect of such deformations is related to their supersymmetry properties.
Since they are realized in terms of changes of the transverse
distribution of the branes, they should preserve an amount of supersymmetry
 and, therefore, this property should
also be manifest in the CFT operators. An analysis in this spirit was performed in
\cite{Prezas:2008ua} for the case of the pointlike system of branes, whose
CFT description involves the linear dilaton theory $\mathbb{R}_\phi$
and the $SU(2)$ Wess--Zumino--Witten (WZW) model.

\no The purpose of the present paper is to analyze aspects of the interplay
between spacetime deformations and the corresponding marginal CFT operators,
 in particular
with respect to their supersymmetry properties, in a third example of a system
with a known exact CFT description. This system is comprised of a set
of NS5- and F1-branes located at the same point in their common transverse space
with four of the Euclidean worldvolume directions of the NS5-branes wrapped on a
4-torus, along which the F1-branes are smeared homogeneously.
In the near-horizon limit it features a constant dilaton and a geometry of the form
$\mathrm{AdS}_3\times S^3\times \mathbb{T}^4$ supported
by appropriate NSNS 3-form fluxes. The exact CFT description is provided by the product of
WZW models
$SL(2,\mathbb{R}) \times SU(2)$ along with four free compact bosons $U(1)^4$
corresponding to $\mathbb{T}^4$.
Notice that the analogue of LST
in this case is a 2-dimensonal CFT residing on the boundary of $\mathrm{\mathrm{AdS}}_3$
which is known to be a deformation of a symmetric orbifold theory. This theory
arises as the infrared limit of the super-Yang--Mills theory
that lives on the common $(1+1)$-dimensional non-compact
worldvolume of the branes. {\em In order to avoid any confusion, we emphasize that
in this paper CFT will always mean the worldsheet theory underlying the
F1-NS5-brane system and not the CFT on the boundary of the  AdS$_3$.}

\no We will start by uncovering the CFT operators dual to some simple deformations
of the brane system. Subsequently, we will perform a 
full-fledged analysis of the supersymmetry
properties of a large class of marginal operators in
the $SL(2,\mathbb{R}) \times SU(2) \times U(1)^4$ theory.
Some of the operators we study have been analyzed in the context
of the ${\rm \mathrm{AdS}}_3/{\rm CFT}_2$
duality starting from \cite{Kutasov:1998zh}.
The most interesting aspect
of this analysis stems from the fact that for backgrounds of this type, i.e.~which
feature  timelike
curved geometries, and as opposed to the case of Minkowski spacetime
\cite{Banks:1988yz},
the existence of spacetime supersymmetry is not tight to
${\cal N}=2$ superconformal (SCFT) invariance on the worldsheet  \cite{Giveon:1998ns}.
Therefore, the set of chiral (or antichiral) primaries, which preserve
automatically the ${\cal N}=2$ SCFT symmetry, provides only a small
subset of the operators that can lead to spacetime supersymmetric deformations.

\no This observation should be compared to what happens
for the first two systems mentioned here. Those comprise only of NS5-branes and consequently time is a non-intracting factor in the sigma model. The  ${\cal N}=2$ superconformal  algebra is realized in a conventional (hermitian)  manner and the set of chiral and antichiral
primaries captures precisely all possible geometric brane deformations
\cite{Fotopoulos:2007rm, Prezas:2008ua}. This no longer holds in the NS5/F1 system under investigation, where  we will uncover, among others, a new class of operators
whose effect on the branes is to perturb the originally homogeneous distribution of the F1-branes inside the NS5-branes.
In other words, if we view the F1-branes as smeared
instantons in the NS5-brane theory, turning on these operators corresponds to
infinitesimal motions in the instanton moduli space.

\no The layout of this paper is as follows. We start in section 2 with a supergravity
analysis of general F1-NS5-brane systems and discuss the exact CFT description
of the pointlike setup as well as certain deformations thereof. In this section
we also present the CFT operators that correspond to the deformations we have performed.
In section 3 we review the construction of the spacetime supercharges of the undeformed
$\mathrm{AdS}_3 \times S^3$ and subsequently we uncover the set of chiral and antichiral
primaries of the worldsheet CFT as well as a large class of spacetime supersymmetry
preserving operators. We discuss  several issues pertaining to the potential
interpretation of those operators in terms of brane deformations. Finally, in the last
section we extend our analysis to a more general class of operators and we provide the
brane description of a class of them that lead to supersymmetric deformations.
In the appendices we have summarized our conventions on the
$SU(2)$ and $SL(2,\mathbb{R})$ WZW models and we have provided the
explicit realization of the ${\cal N}=2$ superconformal algebra employed
in the analysis of the chiral primaries.

\section{F1-NS5-brane configurations}
\label{F1NS5}
\setcounter{equation}{0}
\renewcommand{\theequation}{\thesection.\arabic{equation}}

In this section we study the F1-NS5-brane system from the supergravity and exact
conformal field theory description view points.
In particular, we perform certain symmetric
 perturbations around the point where the exact CFT description
 is known and describe them in terms of WZW primaries and currents of the
associated CFTs.

\subsection{Generic 1/4 supersymmetric configurations}

Our starting point is a 10-dimensional background metric of the form
\be
 ds^2=  H_1^{-1}(-dt^2+dz^2)+
H_5 dx^i dx^i + dy^a dy^a\ , \qq i,a=1,2,3,4\ ,\quad H_{1,5}=H_{1,5}(x)\ ,
\label{generalmetric}
\ee
which for appropriate choices of the functions $H_{1,5}(x)$ represents the gravitational
backreaction of a large collection of F1- and NS5-branes. The worldvolume
of the F1-branes is spanned by  $z^\mu=(t,z),\; \mu=0,1,$ while that of
the NS5-branes by  $z^\mu=(t,z)$ and $y^a$. We will
assume that the 4-dimensional part of the NS5-brane worldvolume parametrized
by $y^a$ is wrapped on a flat 4-torus $\mathbb{T}^4$.
Therefore
both types of branes share a $(1+1)$-dimensional non-compact worldvolume
parametrized by $z^\mu$.

\no The coordinates $x^i$ parametrize the common transverse space
and are non-compact. Notice that since we assume that
$H_1$ depends only on $x^i$
but not on the additional transverse coordinates of the F1-branes $y^a$,
the latter are effectively smeared homogeneously on the 4-torus.
The geometry is supplemented by a dilaton field $\Phi=\Phi(x)$ as well as an NSNS
3-form field strength whose non-vanishing components are $H_{ijk}$
and $H_{tzi}$. These are sourced, respectively, 
by the NS5- and F1-branes.

\no We can choose an orthonormal frame
\be
e^\m = H_1^{-1/2} dz^\m \ ,\qq e^i = H_5^{1/2} dx^i \ ,\qq
\ee
from which we compute the spin connection with non-vanishing elements
\be
\om^{ij}=-\ha H_5^{-1} \del^{[i} H_5 dx^{j]}\ ,\qq \om^{\m i}=-\ha H_1^{-3/2}
H_5^{-1/2} \del^i H_1 dz^\m\ .
\ee
The Killing spinor equations arising by setting to zero the gravitino and
dilatino supersymmetry variations are
\begin{equation}
\label{kns}
\begin{array}{rcl}
\displaystyle{ \del_\m\e + {1\ov 4}\left(\om^{ab}_\m -\ha H_\m{}^{ab}\right)\G_{ab}\e}&=&\displaystyle{0} \ , \crbig
\displaystyle{ \G^\m \del_\m \e -{1\ov 12} H_{\m\n\r}\G^{\m\n\r}\e}&=&\displaystyle{0} \ .
\end{array}
\end{equation}
In addition, we have to satisfy the equations of motion
\begin{equation}
\begin{array}{rcl}
\displaystyle{ R_{\m\n}-{1\ov 4} \left(H^2\right)_{\m\n}+2 D_\m D_\n\Phi}&=&\displaystyle{0} \ , \crbig
\displaystyle{D_\m \left(e^{-2 \Phi}H^\m{}_{\n\r}\right)}&=&\displaystyle{0} \ .
\end{array}
\end{equation}

\no
From the dilatino equation we find  the projections
\be
\G^{tz}\e = \e \ , \qq \G^{1234}\e=-\e\ ,
\label{jfd}
\ee
where the first refers to the common worldvolume directions of the F1- and
NS5-branes, while
the second to the common transverse directions.
These conditions reduce
the amount of preserved supersymmetry to $1/4$ of the original one. Therefore,
for type II
superstring theories we obtain, in the generic case, backgrounds which preserve 8 supersymmetries.

\no
From the gravitino equation we deduce the form of the antisymmetric tensor
field strength (all indices below are curved)
\be
H_{tzi}=\del_i H_1^{-1}\ ,\qq H_{ijk}=\e_{ijk}{}^{l} \del_l H_5\ ,
\label{thh}
\ee
where the index is raised with the flat metric in $\mathbb{R}^4$.
The form of the Killing spinor is $\e=H_1^{-1/4} \e_0$, with $\e_0$ being
a constant spinor subject to the same projections as \eqn{jfd}.
These results, in combination with the dilatino equation, restrict the form
of the dilaton to
\be
e^{-2\Phi}={H_1\ov H_5}\ .
\ee
Finally, the Bianchi identity $dH=0$ requires that
$H_5$ is a harmonic function, while it imposes no condition on $H_1$.
The latter, however, must also be a harmonic function in order that
the equations of motion are satisfied. Therefore we get
\be
\del_i\del^i H_{1,5}=0\ .
\ee
The general solution of those equations is obtained from the (unit-normalized) densities
$\rho_{1,5}(x)$ of F1- and NS5-branes as
\begin{equation}
H_{1,5}(x) = c_{1,5} \int_{\mathbb{R}^4} d{\bf x'} \frac{\rho_{1,5}({\bf x'})}{|{\bf x} - {\bf x'}|^2}\ ,
\end{equation}
where $c_1=  g_{\mathrm{s}}^2 \alpha'^3 N_1 / V_4$ and $ c_5 = \alpha' N_5$. The  numbers
$N_{1,5}$ correspond to the total electric and magnetic NSNS charge.
We focus on the near-horizon region of the branes and thereby
we have dropped the constant term that in principle we could have added
to the harmonic functions.

\subsection{NS5- and F1-branes at a point,
supersymmetry enhancement and exact CFT}

The simplest configuration we can consider is that where
both types of branes reside on the same point  $x^i=0$ in their common transverse
space. Then
\begin{equation}
H_5=\frac{c_5}{r^2}\ , \quad H_1 = \frac{c_1}{r^2}\ ,
\end{equation}
where $r^2=x^i x^i$.

\no
This configuration is particularly interesting for two reasons. First,
the preserved supersymmetry is enhanced to 16 supercharges. This is basically due to the
conformal flatness of the 6-dimensional non-trivial part of the
10-dimensional background and analogous to the supersymmetry enhancement that occurs when we
probe the near horizon region of a D3-brane, where the original 16 supersymmetries
are enhanced to 32.

\no
Second, it is easy to see that the metric  (\ref{generalmetric}) and the antisymmetric-tensor field strength
\eqn{thh}, after a change of coordinates $r= e^{\phi}$ and appropriate rescaling
of $t$ and $z$, take the form
\begin{equation}
\label{ads3xs3}
\begin{array}{rcl}
\displaystyle{ ds^2 }&=&\displaystyle{ \alpha' N_5  \Big(e^{2\phi} (-dt^2 + dz^2) + d\phi^2 +d\Omega_3^2 \Big)+ dy^i dy^i} \ , \crbig
\displaystyle{H}&=&\displaystyle{ 2 \alpha' N_5} \left({\rm Vol}_{\mathrm{AdS}_3} + {\rm Vol}_{S^3}\right) \ ,
\end{array}
\end{equation}
which describes the geometry of $\mathrm{AdS}_3 \times S^3 \times \mathbb{T}^4$, supported by appropriate NSNS fluxes.
Along with the 3-form field strengths, this background admits an exact CFT
description in terms of the WZW models $SL(2,\mathbb{R})\times SU(2)$
and 4 free compact bosons $U(1)^4$ corresponding to $\mathbb{T}^4$.
As is evident from (\ref{ads3xs3}) the level of both cosets is set by $N_5$,
while the number of  F1-branes $N_1$ appears only in the value of the 6-dimensional
string coupling (the constant dilaton)
\begin{equation}\label{gstr}
g_{\mathrm{s}}^2 = \frac{N_5}{N_1}\ .
\end{equation}

\subsection{NS5-branes on a circle and F1-branes at a point}

\subsubsection{Identification of the marginal operators}

For a system of NS5-branes it is known that besides the
pointlike configuration, which admits an exact CFT description
in terms of a linear dilaton $\mathbb{R}_\phi$ theory
and the $SU(2)$ WZW model \cite{Callan:1991dj}, another system that
admits an exact CFT description is that of a circular distribution.
The corresponding CFT, after an appropriate T-duality,  is an orbifold of the
product of the coset models
$SL(2,\mathbb{R})/U(1) \times SU(2)/ U(1)$ for the transverse space, times free bosons for the directions
longitudinal to the NS5-branes \cite{Sfetsos:1998xd}.

\no
An interesting way of thinking about the circular distribution is
as a small deformation of the original pointlike setup. In CFT terms
we can think of the deformed model as arising from
an exactly marginal deformation of the original
$\mathbb{R}_\phi \times SU(2)$ theory \cite{Giveon:1999px, Aharony:2003vk}.
We would like to maintain
this point of view and study the system of NS5-branes on a circle,
this time in the presence of the F1-branes, as a deformation
of the original $SL(2,\mathbb{R})\times SU(2) \times U(1)^4$ theory that
describes the setup where all branes reside at a single point.

\no
Therefore, let us take the
centers of the NS5-branes distributed on an $N_5$-polygon situated
in the plane spanned by $x_3$ and $x_4$ inside the space transverse to the
branes. We have
\be
\vec x_p = r_0(0,0,\cos\phi_p ,\sin\phi_p)\ ,\qq \phi_p= 2\pi{p\ov N_5}\ ,
\quad p=0,1,\ldots , N_5-1\ .
\ee
This distribution of branes preserves an $SO(2)\times \mathbb{Z}_{N_5}$ subgroup
of the original  $SO(4)$ symmetry that is exhibited by the point-like setup.
In the continuum limit the branes are distributed on a ring of
radius $r_0$ situated in the (34)-plane and the symmetry subgroup
becomes continuous, i.e.~$SO(2)\times SO(2)$.
After changing variables as \cite{Sfetsos:1998xd}
\begin{equation}
\label{sfetsosco}
\begin{array}{rclrcl}
\displaystyle{ x_1}&=&\displaystyle{r_0 \sinh\r \cos\th \cos\tau}\ , &\quad
\displaystyle{x_2}&=&\displaystyle{r_0 \sinh\r \cos\th \sin\tau} \ , \crbig
\displaystyle{x_3}&=&\displaystyle{r_0 \cosh\r \sin\th \cos\psi}\ , &\quad
\displaystyle{x_4}&=&\displaystyle{r_0 \cosh\r \sin\th \sin\psi} \ ,
\end{array}
\end{equation}
with ranges
\be
0\leqslant \r<\infty \ , \qq 0\leqslant \th < {\pi\ov 2}\ ,\qq 0\leqslant \psi,\tau< 2\pi\ ,
\ee
we find that the flat metric on $\mathbb{R}^4$ takes the form
\be
dx^idx^i  =   r_0^2\left[
(\sinh^2\r +\cos^2\th)(d\r^2+d\th^2) +\sinh^2\r \cos^2\th\ d\tau^2
+\cosh^2\r \sin^2\th\ d\psi^2\right] \ ,
\ee
while the harmonic function describing the circular distribution of NS5-branes
reads
\be
H_5 =  {c_5\ov \sqrt{(x_1^2+x_2^2+x_3^2+x_4^2+r_0^2)^2-4 r_0^2 (x_3^2+x_4^2)}}
= {c_5/r_0^2\ov \sinh^2 \r + \cos^2\th}\ .
\ee
Instead, since the F1-branes are all located at the origin, we have
\be
H_1={c_1\ov r^2} ={c_1/r_0^2  \ov \sinh^2 \r + \sin^2\th}\ .
\ee
Then, the 6-dimensonal part of the background is\footnote{As in the pointlike
case we can get rid of factors of $g_{\mathrm{s}}, N_1, V_4$ and $r_0$ by
rescaling $t$ and $z$. We will also omit the universal factor
$\alpha' N_5$ to avoid cluttering of the formulas and stick to these conventions
for the rest of the paper. }
\ba
ds^2_6 & = & (\sinh^2\r + \sin^2\th) (-dt^2+dz^2)+
 d\r^2 + d\th^2 +{\tan^2\th d\psi^2 + \tanh^2 \r d\tau^2\ov 1+
\tan^2\th \tanh^2\r} \ ,
\nonumber\\
B_{tz} & = &  \sinh^2\r + \sin^2\th \ ,\qq B_{\tau\psi}\ = \ {1\ov 1+\tan^2\th \tanh^2\r} \ ,
\label{hdg1}
\\
e^{-2 \Phi}& = & {N_1\ov N_5}\ {\sinh^2\r + \cos^2\th \ov \sinh^2\r + \sin^2\th} \ .
\nonumber
\ea

\no
Asymptotically, for $\rho \rightarrow \infty$, this background approaches
$\mathrm{AdS}_3 \times S^3$ which corresponds to the pointlike configuration discussed
in subsection 2.2.
The leading-order corrected metric, due to the circular distribution
of the NS5-branes, is
 \begin{equation}
 ds^2_6 =  d\rho^2+ e^{2 \rho} dx^+ dx^-+ d\Omega^2_3
 + 4e^{-2\rho} (\sin^4 \theta d\psi^2-\cos^4 \theta d\tau^2)
- 2\cos 2\theta dx^+dx^- + \cdots \ ,
 \end{equation}
where we introduced null coordinates $x^\pm =\frac{z\pm t}{2}$.
The corresponding expression for the antisymmetric tensor is
\begin{equation}
B_{\tau\psi} = \cos^2\th + 4 e^{-2\r} \cos^2\th\sin^2\th + \cdots  \ ,\qq
B_{x^+ x^-} =\ha e^{2\r} - \cos\th + \cdots \ .
\end{equation}

\no The first term in the deformation of the metric as well as the deformation of $B_{\tau\psi}$
originate from
  \begin{equation}
e^{-2 \rho} J^3 \bar J^3  \sim  \Phi^{sl}_{0;-1,-1} J^3 \bar J^3\ ,
  \label{su2def}
  \end{equation}
where $\Phi^{sl}_{0;-1,-1}$  is  the normalizable branch of the identity operator in $SL(2,\mathbb{R})$
with conformal dimension 0 and $J_3, \bar J_3$ are the Cartan currents of $SU(2)$
\footnote{The explicit semiclassical expressions for all WZW currents and operators
that we use, can be found in appendices A and B.}.
This is in direct analogy with the deformation of the linear dilaton $\mathbb{R}_\phi$ theory
times the $SU(2)$ WZW model that perturbs a system of pointlike NS5-branes towards a small
circle \cite{Prezas:2008ua}.\footnote{
In that case we have a perturbation of the form $e^{-q\phi}J_3\bar J_3$.
By taking into account the background charge $-q/2$
of a canonically normalized boson, the conformal dimension of $e^{-q \phi}$ is zero.
\label{q}} Notice that in both systems (F1-NS5 and pure NS5), we could use the genuine identity operator instead of the normalizable dimension-zero one, but the corresponding marginal deformation driven by $J_3 \bar J_3$ would not be related to any brane displacement.

\no
The second term in the deformation of the metric as well as the deformation of $B_{x^+x^-}$
resides  in the  $SL(2,\mathbb{R})$ sector of the original CFT.
We can find the corresponding CFT operator by using
the relations in appendices A and B. It reads
\begin{equation}\
-\cos 2 \th  e^{-4\r} \ K^+ \bar K^+ \sim -\Phi^{su}_{1;0,0}  \Phi^{sl}_{1;-2,-2} \ K^+ \bar K^+  \ .
\end{equation}
Notice that due to the fact that the quantum numbers of
$\Phi^{sl}_{1;-2,-2}$ correspond to the highest weight state of the negative
discrete series, its OPEs with $K^+$ and $\bar K^+$ are regular and there is no
normal-ordering ambiguity for the $SL(2,\mathbb{R})$ operators.

\no
To summarize, the deformation of $SL(2,\mathbb{R}) \times SU(2)$
that corresponds to a circular configuration of NS5-branes with the
F1-branes still sitting at a point, takes the form
\begin{equation}
\Phi^{sl}_{0;-1,-1} J^3 \bar J^3 - \Phi^{su}_{1;0,0}   \Phi^{sl}_{1;-2,-2} \; K^+ \bar K^+\ .
\label{eq225}\end{equation}
We can easily see that
both of these operators are marginal. 
 
\subsection{NS5-branes at a point and F1-branes on a circle}

A configuration complementary to the one studied in the previous subsection
is that of NS5-branes residing  at a point with the F1-branes put on a circle.
In this case the appropriate
coordinate system, i.e.~that in which the deformation is manifestly marginal, 
is actually 
\begin{equation}
\label{polco}
\begin{array}{rclrcl}
\displaystyle{x_1}&=&\displaystyle{r   \cos\th \cos\phi}\ , &\quad
\displaystyle{x_2}&=&\displaystyle{ r \cos\th \sin\phi} \ , \crbig
\displaystyle{ x_3}&=&\displaystyle{r \sin\th \cos\tau}\ , &\quad
\displaystyle{x_4}&=&\displaystyle{r \sin\th \sin\tau} \ ,
\end{array}
\end{equation} 
and  the relevant harmonic
functions are now given by
\begin{equation}
H_1 = \frac{1}{ \sqrt{(r^2+1)^2 - 4 r^2 \sin^2 \th}}\ , \qq
H_5= \frac{1}{r^2} \ .
\end{equation}

\no
Performing the same expansion as before yields the leading deformation corresponding
to this background with respect to the unperturbed system where both sets of branes
lie at a point. The deformation contains only $SL(2,\mathbb{R})$ currents and 
reads
\begin{equation}
\label{defone}
\cos 2\th \; \partial x^+ \bar \partial x^- = \Phi^{su}_{1;0,0} \Phi^{sl}_{1;-2,-2} \; K^+ \bar K^+ \ .
\end{equation}
Notice that this differs just by an overall sign from the $SL(2,\mathbb{R})$ deformation that appeared in the previous example. Therefore, if we put both types of
branes on circles of the same size, the total deforming operator will be
$\Phi^{sl}_{0;-1,-1} J^3 \bar J^3$. As we already pointed out,
this is analogous to the  $\mathbb{R}_\phi \times SU(2)$ operator that
deforms the system of NS5-branes
from a point into a small circle \cite{Prezas:2008ua}. Here, we see that this deformation treats
both NS5- and F1-branes on an equal footing
since it corresponds to
putting on a circle {\em of the same radius} both types of branes simultaneously.

 \subsection{Elliptical deformations}

We have seen so far that the $SU(2)$ primary $\Phi^{su}_{1;m,\bar m}$ has appeared with $m=\bar m=0$. The reason for that
is that a circle deformation of either type of branes preserves the $SO(2)$ symmetry associated with
the plane where the deformation takes place. Hence, we expect that a generic planar deformation
will break this isometry and trigger $SU(2)$ primaries with $m,\bar m \neq 0$. To be concrete, let us consider a small elliptical deformation of the F1-branes,
as  described by
 \begin{equation}
 (x^1)^2 + (x^2)^2 = \epsilon^2 \cos^2 \psi\ .
 \end{equation}
 The corresponding deformation of the harmonic function $H_1$ away from
 its point-like limit is, to leading order
 \begin{equation}
 \delta H_1 \sim \frac{1}{r^4} (\cos 2\theta + \cos^2\theta \cos 2\phi) \quad \Longrightarrow\quad
 \delta H_1^{-1} \sim \cos 2\theta + \cos^2\theta \cos 2\phi\ .
 \end{equation}
This can be written in terms of WZW primaries and currents as
  \begin{equation}
(\cos 2\theta +  \cos^2\theta \cos 2\phi)  \ \partial x^+ \bar \partial x^-
\sim  (\Phi^{su}_{1;0,0} +  \Phi^{su}_{1;1,1}   + \Phi^{su}_{1;-1,-1})
   \Phi^{sl}_{1;-2,-2} \; K^+ \bar K^+ \ .
\label{eq233}
\end{equation}
 The first term gives rise to a perturbation that is the same as (\ref{defone}) corresponding
 to the circular deformation of the F1-branes. The other-two terms describe precisely
the breaking of the $U(1)$ symmetry due to the elliptical deformation and come on equal
footing to ensure the reality of the perturbation.

\section{Supersymmetric deformations of  the $SL(2,\mathbb{R}) \times SU(2)$ theory}

We will now proceed with a systematic scan of all operators that can trigger supersymmetric deformations of the original theory. For this purpose, we use worldsheet CFT techniques. As already mentioned in the introduction, 
spacetime supersymmetry does not require in the present framework $\mathcal{N}=2$ superconformal invariance, which turns out to be preserved only for a subset of the
deformations.

\subsection{Spacetime supersymmetry of $\mathrm{AdS}_3 \times S^3$
background}

All the configurations of the previous section preserve at least 1/4 of supersymmetry
and therefore, if embedded in type II superstrings, they should maintain 8
supersymmetries. The special system where
both types of branes are at the same point exhibits actually supersymmetry enhancement
and preserves 16 supersymmetries. This matches the number of spacetime supercharges
constructed in the $\mathrm{AdS}_3 \times S^3$ $\sigma$ model, as we will review shortly
following \cite{Giveon:1998ns}. Subsequently, we would like to establish
that the deformations we found in the previous section preserve 8 supercharges,
in accordance with the analysis of the
Killing spinor equations that we have performed there.

\no
It is standard lore in string theory that spacetime supersymmetry
is tied to the existence of extended worldsheet supersymmetry. However,
the fact that we deal here with a curved timelike background, due to the
$\mathrm{AdS}_3$ factor in the metric, invalidates the usual argument due to \cite{Banks:1988yz},
which refers to a Minkowski spacetime,
and one has to follow a different procedure. The approach
of  \cite{Giveon:1998ns} was to construct the spacetime supercharges
directly, i.e.~without employing the underlying ${\cal N}=2$ supeconformal symmetry,
and explicitly verify their BRST invariance. We will proceed here in a similar fashion and further discuss the appearance and the role of an ${\cal N}=2$ superconformal algebra in Sec. \ref{CP}.


\no
In order to construct the spacetime supercharges we should first bosonize the
fermions of the theory.
The fermions
$\psi^\pm, \chi^\pm$ and $\chi^3, \psi^3$ are bosonized
in terms of three canonically normalized scalars fields. In order to capture all fermions, including the partners of the $\mathbb{T}^4$, we will introduce
five bosons
 $H_i, i=1,\ldots ,5$,
obeying
\begin{equation}
H_{i}(z) H_{j}(w) =  -\delta_{ij}\log(z-w) \ ,i,j = 1,2,\ldots ,5\ .
\end{equation}
Recall that for a scalar field with canonical normalization we have
\begin{equation}
e^{i a H(z)} e^{i b H(w)} = (z-w)^{ab} e^{i [a H(z)+ b H(w)]} \ ,
\end{equation}
where normal ordering is implied.
Then, we have
\begin{eqnarray}
\psi^\pm = e^{\pm i H_1}\ , \quad \quad \chi^\pm = e^{\pm i H_2} , \quad\quad
\psi^3 = \frac{e^{i H_3} + e^{-i H_3}}{\sqrt{2}}\ , \quad\quad
\chi^3 = \frac{e^{i H_3} - e^{-i H_3}}{\sqrt{2}}\ .
\end{eqnarray}
Correspondingly, we have the currents
\begin{equation}
\psi^+ \psi^- = i \partial H_1\ , \quad
\chi^+ \chi^- = i \partial H_2\ , \quad
\psi^3 \chi^3 = - i \partial H_3\ .
\end{equation}
The expression for $\chi^3$ reflects  the fact that its norm is negative.
The fermions $\lambda^a, a=1,\ldots,4$ of the $\mathbb{T}^4$
are bosonized in a standard fashion as
\begin{equation}
\hat \lambda^{\pm}:=
\frac{1}{\sqrt{2}}(\lambda^1\pm i \lambda^2) = e^{\pm i H_4}, \qq
\tilde \lambda^{\pm} := \frac{1}{\sqrt{2}}(\lambda^3\pm i \lambda^4) =  e^{\pm i H_5}\ .
\end{equation}
Notice that $H^\dagger_{1,2,4,5}= H_{1,2,4,5}$ while $H^\dagger_3=-H_3$.

\no
The supercharges take the usual form
\begin{equation}
Q=\oint dz e^{-\frac{\varphi}{2}} e^{\frac{i}{2} \sum_{i=1}^5 \epsilon_i H_i}\ ,
\end{equation}
with $\epsilon_i$ being $\pm 1$ and $\varphi$ being the bosonized superghost.
The allowed values of $\epsilon_i$ are constrained due to the
requirement of mutual-locality, which demands
\begin{equation}
\prod_{i=1}^5 \epsilon_i = 1\ ,
\end{equation}
and BRST invariance, which further dictates
\begin{equation}
\prod_{i=1}^3 \epsilon_i = -1\ .
\label{supercon}
\end{equation}
It is fairly straightforward to see why the first condition is necessary.

\no
The second condition comes out as follows. The BRST charge contains
a term $Q_{\rm BRST} =  \cdots + \gamma G^1+\cdots$, where $\gamma$ is
one of the superghosts and the ${\cal N}=1$ supercurrent $G^1=\frac{1}{\sqrt{2}}
(G^+ + G^-)$ contains the cubic terms
\begin{equation}
G^1_{\rm 3-Fermi} = \psi^+\psi^-\psi^3 -\chi^+ \chi^-\chi^3
\sim (\partial H_1 - \partial H_2) e^{i H_3} +  (\partial H_1 + \partial H_2) e^{-i H_3}\ ,
\end{equation}
as can be found from the realization \eqn{emtpart}.
These terms  can give poles of  order ${\cal O}(z^{-3/2})$
and  ${\cal O}(z^{-1/2})$ in their OPE with the supercharges.
Since the OPE of the superghost $\gamma$ with $e^{-\frac{\varphi}{2}}$
is of order ${\cal O}(z^{1/2})$, the only potential problem comes from the
${\cal O}(z^{-3/2})$ poles, which therefore should cancel out.
Explicitly, we find that the OPE
\begin{equation}
G^1_{\rm 3-Fermi}(z)\  e^{\frac{i}{2}[\epsilon_1 H_1(w) + \epsilon_2 H_2(w) + \epsilon_3 H_3(w)]}\ ,
\end{equation}
is proportional to
\begin{equation}
\frac{\epsilon_1 -\epsilon_2 }{z-w} (z-w)^{\frac{\epsilon_3}{2}}+
\frac{\epsilon_1 +\epsilon_2 }{z-w} (z-w)^{-\frac{\epsilon_3}{2}}\ .
\end{equation}
Therefore, if $\epsilon_3=1$ we need $\epsilon_1+\epsilon_2=0$ to cancel
the  ${\cal O}(z^{-3/2})$ pole  from the second term and vice versa if
$\epsilon_3=-1$, i.e.~we obtain condition
(\ref{supercon}).

\no
Summarizing, the allowed supercharges are
\begin{equation}
\label{spacesusy}
\begin{array}{rcl}
\displaystyle{Q_{1\pm} }&=&\displaystyle{  e^{\frac{i}{2}[-H_1-H_2-H_3\pm (H_4-H_5)]}} \ , \crbig
\displaystyle{Q_{2\pm} }&=&\displaystyle{  e^{\frac{i}{2}[-H_1+H_2+H_3\pm (H_4-H_5)]}} \ , \crbig
\displaystyle{ Q_{3\pm}}&=&\displaystyle{e^{\frac{i}{2}[H_1+H_2-H_3\pm (H_4-H_5)]}} \ , \crbig
\displaystyle{Q_{4\pm }}&=&\displaystyle{e^{\frac{i}{2}[H_1-H_2+H_3\pm (H_4-H_5)]}} \ .
\end{array}
\end{equation}
These are  8 supercharges and
along with the contribution from the antiholomorphic sector we obtain in
total 16 supercharges, which matches the number of supersymmetries preserved
by the dual brane system.

\subsection{Chiral primaries} \label{CP}

Before proceeding with the analysis of the various supersymmetric deformations and preserved spacetime supercharges (Sec. \ref{ssd}), we would like to pause and discuss the advertized superconformal symmetry.

\no In a theory with  ${\cal N}=2$ superconformal symmetry
 one can obtain a class of worldsheet supersymmetry-preserving marginal deformations
 by considering the chiral (and antichiral) primary operators.
 Since, however, for the backgrounds of interest the existence of spacetime
 supersymmetry is not tied to the ${\cal N}=2$ worldsheet supersymmetry,
 one should not restrict to chiral primaries. As we will see below,
the deformations originating from chiral primaries 
are indeed a small subset of the class of deformations
 preserving spacetime supersymmetry.

\no 
The reader might be puzzled by the above statement, referring to an $\mathcal{N}=2$ superconformal algebra, which is not expected to be realized in Lorentzian backgrounds. In the $\sigma$ model under consideration, however, a non-hermitian realization of such an algebra is available and displayed in appendix C. It can be understood as follows: The non-trivial part of the worldsheet theory is the factor $SL(2,\mathbb{R})\times SU(2)$, which can be further decomposed  as $\frac{SL(2,\mathbb{R})}{U(1)}\times \frac{SU(2)}{U(1)} \times U(1)\times \mathbb{R}$. The coset factor 
  $\frac{SL(2,\mathbb{R})}{U(1)}\times \frac{SU(2)}{U(1)} $ provides a genuine  $\mathcal{N}=2$ (even $\mathcal{N}=4$) superconformal algebra -- the one present e.g. in the circular NS5-brane distribution -- whereas the lightcone factor $U(1) \times \mathbb{R}$  is presumably responsible for the lack of hermiticity. For our purposes, it is obviously natural to use the primaries and currents  of the $SL(2,\mathbb{R})$ and $SU(2)$ WZW models. Any further reference to the $\mathcal{N}=2$ algebra should be understood in those terms. 
  
 \no 
Returning to our analysis we would like  to use the chiral primary operators as supersymmetric
 seeds for  marginal deformations so that we will focus on those that
 have conformal dimension $h=1/2$. Subsequently, their $R$-charge
 should be $Q=\pm 1$.
A quite broad class of operators with $h=1/2$ has the following form\footnote{
We will focus mostly on normalizable operators in the $SL(2,\mathbb{R})$ model,
 since these correspond
to deformations of the brane system. In other words by $\Phi^{sl}_{j;m}$
we mean the normalizable version of the operator with
conformal weight $\D=-j(j+1)/k$. Recall that
to each such  conformal weight in $SL(2,\mathbb{R})$
there are associated two values of $j$ related by reflection $j \leftrightarrow -j-1$.
The two values correspond to the normalizable and non-normalizable branch
of the corresponding operator.
For instance, the non-normalizable identity operator with $\D=0$ has
$j=-1, m=0$ and is annihilated by all $SL(2,\mathbb{R})$
currents, in other  words $\Phi^{sl}_{-1;0}\equiv 1$.
This is the
analogue of $1$ in the linear dilaton theory.
Instead,
the operator with $j=0$ is its normalizable version and the edge states of the
two discrete representations with $m=\pm 1$
correspond to $e^{-q\phi}$ in the linear dilaton theory (see also comment in footnote \ref{q}).}
\begin{equation}
 \Phi^{su}_{j;m} \Phi^{sl}_{j;m'} {\cal Y}\ ,
 \label{susly}
\end{equation}
where $\Phi^{su}_{j;m}, \Phi^{sl}_{j;m'}$
are affine primaries of the bosonic subalgebra of the full affine algebra of the
super-WZW models and  ${\cal Y}$ is any of the fermions of the theory.
In this section we will  actually
restrict our analysis  to the case where ${\cal Y}$ is a fermion in the $SU(2)$
or the $SL(2,\mathbb{R})$ WZW models, since these operators are most relevant
for the applications we have in mind, and we will consider the additional case where ${\cal Y}$
is a fermion from $\mathbb{T}^4$ in the next section.
As usual we have
suppressed the antiholomorphic indices in order to avoid unnecessary cluttering of the formulas.
We hasten to add  that for non-unitary CFTs
 the relation $h=\frac{Q}{2}$ is a necessary but not a sufficient condition for an operator
to be chiral primary. Therefore, we can use it to restrict the possibilities, but
we should still check explicitly if the operators we obtain are actual chiral primaries.

\no
We start by noticing that under the $U(1)$ $R$-current \eqn{emtcur}, $\psi^3\pm\chi^3$ have charges $Q=\mp1$.
Instead, the other fermions have also a contribution from the fermionic part
inside $J^3_{\mathrm{T}}$ or $K^3_{\mathrm{T}}$. Therefore, if we use these fermions
we should appropriately adjust $m$ and $m'$ in order to
have vanishing $J^3_{\mathrm{T}}+K^3_{\mathrm{T}}$ charge and just obtain $Q=\pm1$
from the other fermionic terms. The same is true of course
when the fermion is $\psi^3\pm\chi^3$, where we should
ensure that $m+m'=0$.

\no
Therefore, we conclude that we have the following three classes
of potential chiral primary operators
\begin{eqnarray}
&& \Phi^{su}_{j;m} \Phi^{sl}_{j;m'} \chi^+\ , \qq\qq\ m+m'+1 =0\ ,
\nonumber \\
&& \Phi^{su}_{j;m} \Phi^{sl}_{j;m'} \psi^-\ , \qq\qq\ m+m'-1 = 0\ , \\
&& \Phi^{su}_{j;m} \Phi^{sl}_{j;m'} (\psi^3-\chi^3)\ ,\qq  m+m' = 0
\ .\nonumber
\end{eqnarray}
Similarly we have a complementary set of potential  antichiral operators
with the appropriate fermions.
So far these results do not depend on the particular values of $m$ and $m'$
or on the branch we choose for the $SL(2,\mathbb{R})$ primary.
However,  checking explicitly the chirality of these operators by computing
their OPEs with $G^+$, reveals that, like the situation encountered in
\cite{Prezas:2008ua}, only for specific charges $m$ and $m'$ and specific branches
these operators are actually chiral primary.

\no
From the first two classes we find that only
$\Phi^{su}_{j;j} \Phi^{sl}_{j;-j-1} \chi^+$ and
$\Phi^{su}_{j;-j} \Phi^{sl}_{j;j+1} \psi^-$ are chiral primary.
It is worth noticing that, had we considered the non-normalizable primary of $SL(2,\mathbb{R})$
we would have found that the operator fails to be either primary or chiral.
The purely bosonic pieces of the corresponding deformations are $K^+  \Phi^{su}_{j;j} \Phi^{sl}_{j;-j-1}$ and
$J^- \Phi^{su}_{j;-j} \Phi^{sl}_{j;j+1}$ respectively and
they have vanishing $R$-charge as expected.
Obviously a similar story holds for the antichiral operators which read
 $\Phi^{su}_{j;-j} \Phi^{sl}_{j;j+1} \chi^-$ and
$\Phi^{su}_{j;j} \Phi^{sl}_{j;-j-1} \psi^+$, and which give rise to the deformations
 $K^- \Phi^{su}_{j;-j} \Phi^{sl}_{j;j+1} $ and
$J^+ \Phi^{su}_{j;j} \Phi^{sl}_{j;-j-1} $.
There are no normal-ordering ambiguities since the primaries of the
$SU(2)$ and $SL(2,\mathbb{R})$ WZW theories that appear correspond to edge states of the spin $j$
representations and are annihilated by the associated, with the perturbation current, operators.
From the third class only the operator
$\psi^3 - \chi^3$ is chiral primary
 and leads to the deformation $J^3-K^3$. Notice that this last operator
 is actually non-normalizable in $SL(2,\mathbb{R})$.

\no
 To summarize, the chiral primaries of the theory are
\begin{equation}
\Phi^{su}_{j;j} \Phi^{sl}_{j;-j-1} \chi^+, \qq
\Phi^{su}_{j;-j} \Phi^{sl}_{j;j+1} \psi^-,\qq
\psi^3 - \chi^3\
\end{equation}
and similarly  the antichiral primaries are
\begin{equation}
\Phi^{su}_{j;-j} \Phi^{sl}_{j;j+1} \chi^-, \qq
\Phi^{su}_{j;j} \Phi^{sl}_{j;-j-1} \psi^+,\qq
\psi^3 + \chi^3\ .
\end{equation}

\subsection{Spacetime supersymmetric deformations} \label{ssd}

We note that some of the deforming operators uncovered in the section \ref{F1NS5}
do not originate from the chiral primaries found above. For instance, consider
$ \Phi^{su}_{1;0,0} \Phi^{sl}_{1;-2,-2} \; K^+ \bar K^+$ in  \eqn{eq225}
coming from the seed operator $ \Phi^{su}_{1;0,0} \Phi^{sl}_{1;-2,-2} \; \chi^+ \bar \chi^+$.
The latter does not have the proper $SU(2)$ charge to be a chiral primary.
Since, however, the
deformations arising from chiral primaries are guaranteed
to preserve only the ${\cal N}=2$ worldsheet supersymmetry,
but not spacetime supersymmetry and, in any case, the ${\cal N}=2$
does not seem to be tied to the existence of spacetime supersymmetry,
we should check directly how many of the original supercharges
are conserved by a very general class of deformations. Our findings will  be
in full consistency with the results of the section \ref{F1NS5}, which were based
on supergravity.

 \paragraph{Fermions in the $SL(2,\mathbb{R})$:}

We will consider a general ansatz for a
seed operator of the type studied in \cite{Kutasov:1998zh},
with form
\be
A \Phi^{su}_{j;n} \Phi^{sl}_{j;m+1} \chi^- + B \Phi^{su}_{j;n} \Phi^{sl}_{j;m} \chi^3
+ C  \Phi^{su}_{j;n} \Phi^{sl}_{j;m-1} \chi^+\ .
\ee
Notice that we will restrict ourselves only to NS sector operators.
For certain values of
$A,B,C$, corresponding to Clebsch--Gordan coefficients,
this operator belongs to an irreducible representation with spin $j+1$
of the $SL(2,\mathbb{R})$ generated by the total currents $K^i_{\mathrm{T}}$.
Acting on it with the ${\cal N}=1$ supercurrent $G^1$ and collecting the
residues of the first order pole yields the actual deformation. The latter
consists of the purely bosonic piece
\begin{equation}
A  \Phi^{su}_{j;n} \Phi^{sl}_{j;m+1} K^- + \sqrt{2}  B \Phi^{su}_{j;n} \Phi^{sl}_{j;m} K^3+
 C \Phi^{su}_{j;n} \Phi^{sl}_{j;m-1} K^+\ ,
\end{equation}
as well fermion bilinears.

\no
It is obvious that all spacetime supercharges (\ref{spacesusy})
commute with the purely bosonic
piece of the deformation and potential obstructions result from
the fermion bilinear pieces. Grouping the latter  according to the
bosonic primaries they contain, since different primaries do not interfere with each other,
we have:
\ba
\label{chilbil}
&& (j+n)  C  \Phi^{su}_{j;n-1} \Phi^{sl}_{j;m-1} \psi^+ \chi^+ \ ,
\nonumber\\
&&
 \Phi^{su}_{j;n} \Phi^{sl}_{j;m-1} \Big( \big(B( -1 -  j +  m) + \sqrt{2} C m \big) \chi^+ \chi^3 +
 \sqrt{2} C n \psi^3 \chi^+\Big)\ ,
\nonumber\\
 &&
(j-n) C  \Phi^{su}_{j;n+1} \Phi^{sl}_{j;m-1} \psi^- \chi^+ \ ,
\nonumber\\
&&
(j+n) B   \Phi^{su}_{j;n-1} \Phi^{sl}_{j;m} \psi^+ \chi^3\ ,
\nonumber\\
&&
\Phi^{su}_{j;n} \Phi^{sl}_{j;m}  \Big(
\big(\sqrt{2} B - A( j-m) - C( j  +m)\big) \chi^+\chi^- + \sqrt{2} B n \psi^3 \chi^3 \Big)\ ,
\\
&&
(j-n) B \Phi^{su}_{j;n+1} \Phi^{sl}_{j;m} \psi^- \chi^3\ ,
\nonumber\\
&&
(j+n) A \Phi^{su}_{j;n-1} \Phi^{sl}_{j;m+1} \psi^+ \chi^-\ ,
\nonumber\\
&&
 \Phi^{su}_{j;n} \Phi^{sl}_{j;m+1} \Big(
 \big(B ( 1+ j +m) + \sqrt{2} A m \big) \chi^- \chi^3 +
 \sqrt{2} A n \psi^3 \chi^-\Big)\ ,
\nonumber\\
&&
(j-n) A  \Phi^{su}_{j;n+1} \Phi^{sl}_{j;m+1} \psi^- \chi^-\ .
\nonumber
\ea
The term in the 5th line is a current and its action on any supercharge has always a pole
since all supercharges contain $H_2$ and $H_3$. The condition it leads to
is
\begin{equation}
\big(\sqrt{2} B - A (j-m) - C( j + m)\big) \epsilon_2 - \sqrt{2} B n
\epsilon_3 =0\ .
\label{hfk2}
\end{equation}

\no
In total we have 15 fermion bilinears.
We present in the table
below the result of the action of the fermion bilinears on the supercharges\footnote{In the table we suppress the indices $\pm$ from the supercharges to avoid cluttering.} where a tick
means that the supercharge commutes with
the bilinear.
We have excluded the current terms  $\psi^+ \psi^-$, $\chi^+ \chi^-$ and $ \psi^3 \chi^3$ since
they do  not commute with any supercharge.

\begin{center}
\begin{tabular}{ | c || c | c |  c |c |c |c |c |c |c |c | }
\hline
   & $\psi^+ \chi^+$ & $\chi^+ (\psi^3,\chi^3)$  & $\psi^- \chi^+$
  & $\psi^+ (\psi^3,\chi^3)$ & $\psi^- (\psi^3,\chi^3)$ & $\psi^+ \chi^-$
& $\chi^- (\psi^3,\chi^3)$ & $\psi^- \chi^-$    \\
  \hline\hline
  $Q_{1}$ &  $Q_3$ & $Q_2$  &$\surd$  & $Q_3$   & $\surd$ &$\surd$  & $\surd$    &$\surd$    \\
  \hline
  $Q_{2}$ &  $\surd$& $\surd$  & $\surd$ &  $(Q_3,-Q_3)$ & $\surd$ & $Q_4$ & $(Q_1,-Q_1)$   & $\surd$    \\
  \hline
   $Q_{3}$ &$\surd$  &  $\surd$  &  $\surd$& $\surd$ & $Q_2$  & $\surd$ &  $Q_4$  & $Q_1$ \\
  \hline
  $Q_{4}$ & $\surd$ & $(Q_3,-Q_3)$ & $Q_2$   &$\surd$  &  $(Q_1,-Q_1)$ &  $\surd$& $\surd$    & $\surd$    \\
  \hline
\end{tabular}
\end{center}

\no
Let us now analyze the conditions for preserving at least 4 supercharges,
for instance $Q_{2\pm}$ and $Q_{3\pm}$.
Then we get the six conditions
\be
(j\pm n) A = 0 \ ,\quad (j\pm n) B =0 \ ,\quad (1+j+m) B + \sqrt{2} A m = 0 \ ,
\quad n A= 0 \ ,
\ee
plus two more from \eqn{hfk2} corresponding to $Q_{2\pm}$ (with $\e_2=\e_3=1$)
and $Q_{3\pm}$ (with $\e_2=-\e_3=1$).
Except for the case $j=n=0$ the only solution is
$A=B=0$.
From the current condition $C(j+m) =0$  and
since $C\neq 0$ with obtain eventually $m=-j$. Therefore
the seed operator that leads to a deformation preserving 4 supercharges,
from the holomorphic  sector, is $ \Phi^{su}_{j;n} \Phi^{sl}_{j;-j-1} \chi^+$.
Similarly the operator
$ \Phi^{su}_{j;n} \Phi^{sl}_{j;j+1} \chi^-$ preserves the complementary set
of supercharges $Q_{1\pm}$ and $Q_{4\pm}$. Furthermore, it is straightforward to
check that there are no other combinations of supercharges that can be
preserved except for the two ones above.

 \no
 If $j=n=0$ we find that the supercharges  $Q_{2\pm}$ and $Q_{3\pm}$ are preserved
 provided that
 \begin{equation}
\begin{array}{rcl}
\displaystyle{B(1+m)+\sqrt{2} A m }&=&\displaystyle{0} \ , \crbig
\displaystyle{ \sqrt{2} B + m (A-C) }&=&\displaystyle{0} \ .
\end{array}
\end{equation}
 However these two conditions (along with $n=0$) imply
that the 2nd term in (\ref{chilbil}) has vanishing coefficient and therefore
  $Q_{1\pm}$ and $Q_{4\pm}$ are also preserved! If $m\neq 0$ the general
  solution of that system yields the deforming operator (up to an overall
  multiplicative constant)
  \begin{equation}
  (m+1)  \Phi^{sl}_{0;m+1} K^- - 2  m  \Phi^{sl}_{0;m} K^3+
  (m-1) \Phi^{sl}_{0;m-1} K^+\ ,
\end{equation}
while for $m=0$ we have the deforming operator
\begin{equation}
A  \Phi^{sl}_{0;1} K^-  + C  \Phi^{sl}_{0;-1} K^+\ .
\end{equation}

 \paragraph{Fermions in the $SU(2)$:}

Let us consider now operators of the form
\be
A \Phi^{su}_{j;n+1} \Phi^{sl}_{j;m} \psi^- + B \Phi^{su}_{j;n} \Phi^{sl}_{j;m} \psi^3
+ C  \Phi^{su}_{j;n-1} \Phi^{sl}_{j;m} \psi^+\ .
\ee
The purely bosonic piece of the deformation induced by this operator reads
\begin{equation}
 A \Phi^{su}_{j;n+1} \Phi^{sl}_{j;m} J^- + \sqrt{2} B \Phi^{su}_{j;n} \Phi^{sl}_{j;m} J^3
+ C  \Phi^{su}_{j;n-1} \Phi^{sl}_{j;m} J^+\ .
\end{equation}
The fermion bilinear terms are grouped again according to the
bosonic primaries as follows:
\ba
&& (j+1-m)  C  \Phi^{su}_{j;n-1} \Phi^{sl}_{j;m-1} \psi^+ \chi^+\ ,
\nonumber\\
&&
(j+1-m ) B  \Phi^{su}_{j;n} \Phi^{sl}_{j;m-1} \psi^3 \chi^+\ ,
\nonumber\\
&&
(j+1-m) A  \Phi^{su}_{j;n+1} \Phi^{sl}_{j;m-1} \psi^- \chi^+\ ,\nonumber\\
&&
 \Phi^{su}_{j;n-1} \Phi^{sl}_{j;m}\Big(
 \sqrt{2} C m \psi^+ \chi^3 + \big(\sqrt{2}C+ B (j-n)\big) \psi^+ \psi^3\Big)\ ,\nonumber\\
&&
 \Phi^{su}_{j;n} \Phi^{sl}_{j;m}\Big(\big(\sqrt{2} B +A(1+j+n)-C(1+j-n)\big)\psi^+ \psi^-
 +\sqrt{2} B m \psi^3 \chi^3\Big)\ ,\\
&&
  \Phi^{su}_{j;n+1} \Phi^{sl}_{j;m}\Big(\sqrt{2} A m \psi^- \chi^3 +\big( B(j-n) -\sqrt{2}A n
  \big) \psi^- \psi^3\Big)\ ,\nonumber\\
&&
(j+1+m)  C  \Phi^{su}_{j;n-1} \Phi^{sl}_{j;m+1} \psi^+ \chi^-\ ,\nonumber\\
&&
(j+1+m) B \Phi^{su}_{j;n} \Phi^{sl}_{j;m+1} \psi^3 \chi^-\ ,\nonumber\\
&&
(j+1+m) A  \Phi^{su}_{j;n+1} \Phi^{sl}_{j;m+1} \psi^- \chi^-\ .\
\nonumber
\ea
It is straightforward to check that
$ \Phi^{su}_{j;j} \Phi^{sl}_{j;m} \psi^+$ and $ \Phi^{su}_{j;-j} \Phi^{sl}_{j;m} \psi^-$
preserve half of the original supercharges, that is ($Q_{3\pm},Q_{4\pm}$)
and ($Q_{1\pm},Q_{2\pm}$), respectively. Furthermore, the
operators $\Phi^{sl}_{0;\pm 1}\psi^3$ preserve the
supercharges $Q_{1,\pm}, Q_{4\pm}$  for $m=1$ and
 $Q_{2,\pm}, Q_{3\pm}$ for $m=-1$. No other combinations of supercharges
 can be preserved.
 We should emphasize that
in the general analysis above we take the $SL(2,\mathbb{R})$ primaries in the normalizable branch.  
  
\no
All these operators contain the chiral/antichiral primaries  found
previously but in general there are by far more 1/2-BPS operators.
It can be checked that these extra operators lead to deformations
that do not preserve the original ${\cal N}=2$ SCFT symmetry \footnote{
Notice that in order to have an ${\cal N}=2$ preserving deformation
it is sufficient but not necessary that the seed operator is
chiral or antichiral primary.}.
As a rule, the charge of the WZW
primary that comes from the same model as the fermion is fixed while
the other primary has arbitrary charge.  We should also mention that although
the marginal deformations originating from chiral primaries can be argued to
be exactly marginal, this is not possible for the deformations coming
from the above operators (although that does not necessarily imply that
these deformations are not exactly marginal).
Note also that our operators do not match, in general, the operators of \cite{Kutasov:1998zh},
where $A,B,C$ are fixed in terms of Clebsch--Gordan coefficients,
except when one of the states is at the boundary of the representation space and
two out of the three Clebsch--Gordan coefficients vanish.

 \paragraph{A mixed operator:}

Let us  finally check the operator $\Phi^{sl}_{j;m}
(\psi^3 - \chi^3)$. It is chiral primary for $j=-1,m=0$
(when $\Phi^{sl}_{-1;0} \equiv 1$) and then corresponds to a non-normalizable deformation.
It makes also sense as a seed operator if  $j=0$ and $m=-1$ so that it gives rise
to a marginal normalizable deformation. The bosonic piece of this deformation reads
\begin{equation}
\Phi^{sl}_{j;m}(J^3-K^3)\ ,
\end{equation}
while the fermionic one is given by the sum of the following terms
\ba
&&
(1+j-m) \Phi^{sl}_ {j;m-1} (\psi^3 \chi^+ + \chi^+ \chi^3)\ , \nonumber\\
&& \Phi^{sl}_{j;m} (\chi^+ \chi^- -m \psi^3 \chi^3 - \psi^+ \psi^-)\ ,
\\
&& (1+j+m)  \Phi^{sl}_ {j;m+1}( \psi^3 \chi^-+ \chi^- \chi^3)\ .
\nonumber
\label{eq327}
\ea
The term in the second line implies that $\e_2 + m\e_3 - \e_1 = 0$, which is not possible to satisfy
for $m=-1$. For $m=0$, this condition becomes $\e_1=\e_2$ which is satisfied only for the supercharges
$Q_{1\pm}$ and $Q_{3\pm}$. Then, using also $j=-1$ we see that the terms in the  first and third lines
in \eqn{eq327} are vanishing as well.
Hence, we have a 1/2 BPS deformation, which however is non-normalizable.
One could further consider more general combinations of operators
with fermions from both WZW models, however it turns out that
they do not lead to other supersymmetric operators besides the one
we found above.

 \paragraph{Summary:}

To summarize, we have found the following classes of seed operators
that yield 1/2 BPS deformations in spacetime:
\begin{equation}
\Phi^{su}_{j;n} \Phi^{sl}_{j;\mp j \mp 1} \chi^\pm\ , \quad j\neq 0\ ,  \qq
\Phi^{su}_{j;\pm j} \Phi^{sl}_{j;m} \psi^\pm\ , \qq  \Phi^{sl}_{0;\pm 1}\psi^3\ ,
\qq \psi^3 - \chi^3\ .
\end{equation}
In addition, the following operators yields deformations that do not break
any supersymmetry:
 \begin{equation}
 \label{fullsusy}
 (m+1) \Phi^{sl}_{0;m+1} \chi^- - \sqrt{2} m \Phi^{sl}_{0;m} \chi^3+\
 (m-1) \Phi^{sl}_{0;m-1} \chi^+\ , \quad m \neq 0
 \end{equation}
 and
 \begin{equation}
A  \Phi^{sl}_{0;1} \chi^-  + C  \Phi^{sl}_{0;-1} \chi^+\ .
\end{equation}
Finally, let us  mention that none of the operators we have studied so far can preserve {\em only}
1/4 of the original
supersymmetry.

\subsection{Brane interpretation and comments}

All geometric deformations of the pointlike
brane system are captured by the ansatz (\ref{generalmetric}) with
the functions $H_1$ and $H_5$ depending on the common transverse space, i.e.~on the
radial coordinate $\rho$ as well as on the $SU(2)$ coordinates. Therefore, from the
set of spacetime supersymmetric operators we uncovered in the previous subsection
only a subclass can be given an interpretation in terms of a deformed brane
system. This is the subclass whose $SL(2,\mathbb{R})$ primary
depends only on $\rho$ and which involves only the $K^+, \bar K^+$ currents.
Otherwise it is easy to see, using the formulas from the appendix B, that
the deformation will depend also on the coordinates $x^\pm$, therefore
loosing its brane description. Hence, 
\textbf{the operators that could a priori}
correspond to geometric deformations of the brane system are
\begin{equation}
\label{opbr}
\Phi^{su}_{j;n,\bar n} \Phi^{sl}_{j;- j - 1,-j-1} \chi^+ \bar \chi^+\ , j\neq 0, \qq
\Phi^{su}_{j;\pm j,\pm j } \Phi^{sl}_{j;-j-1,-j-1} \psi^\pm \bar \psi^\pm\ ,
\qq  \Phi^{sl}_{0;- 1,-1}\psi^3 \bar \psi^3\ ,
\end{equation}
where we reinstalled the anti-holomorphic indices for concreteness.

\no
Notice that the only maximally supersymmetric operator that has the right form to yield a brane deformation is $ \Phi^{sl}_{0;-1,-1} \chi^+ \bar \chi^+
\sim e^{2\rho} \partial x^+ \bar \partial x^-$ and therefore it trivially
corresponds to an overall rescaling of the coordinates $x^\pm$. This is
consistent as there are no deformations of the original F1-NS5-brane system
that preserve its total supersymmetry. The rest of the operators
that preserve the full supersymmetry correspond to
diffeomorphisms of the AdS$_3$ metric, as can be verified by computing
the scalar curvature of the deformed metric,  and therefore they have a trivial
physical effect.

\no
We see now  that $\Phi^{sl}_{0;-1,-1} J^3 \bar J^3$ and
$\Phi^{su}_{1;0,0} \Phi^{sl}_{1;-2,-2} \; K^+ \bar K^+ $, which appear when we put the branes on circles,
are accounted for by the third and first operators of the above
list, respectively. It is also important that these two classes of operators preserve
the same set of supercharges, that is $Q_{2\pm}$ and $Q_{3\pm}$,
so that the combined deformation is still
supersymmetric as it should.
The same is true for the operators $ \Phi^{su}_{1;1,1}
  \Phi^{sl}_{1;-2,-2} \; K^+ \bar K^+ $ and $\Phi^{su}_{1;-1,-1} \Phi^{sl}_{1;-2,-2} \; K^+ \bar K^+ $
that describe an elliptical deformation of the F1-branes. We notice now that
from the supergravity
point of view all deformations preserve the same set of supercharges, since
the form of the Killing spinors is not related to the actual expressions for
the harmonic functions $H_1$ and $H_5$, and therefore the operators that
yield brane deformations should be only those commuting with the supercharges
$Q_{2\pm}$ and $Q_{3\pm}$ that are preserved by the circular and elliptical
deformation. Therefore out of  (\ref{opbr}) we should further restrict only
to the operators
\begin{equation}
\Phi^{su}_{j;n,\bar n} \Phi^{sl}_{j;- j - 1,-j-1} \chi^+ \bar \chi^+\ ,\quad  j \neq 0\ , \qq
 \Phi^{sl}_{0;- 1,-1}\psi^3 \bar \psi^3\ .
\end{equation}

 \no
 It is a bit surprising that  $\Phi^{su}_{j;\pm j,\pm j } \Phi^{sl}_{j;-j-1,-j-1} \psi^\pm \bar \psi^\pm$ have to be excluded since similar operators in \cite{Prezas:2008ua},
 containing linear dilaton vertex operators instead of $SL(2,\mathbb{R})$
 primaries, where argued to account for the geometric deformations of the
 pointlike NS5-brane system (along with the analogue of   $ \Phi^{sl}_{0;- 1,-1}\psi^3 \bar \psi^3$ that describes the circular deformation). However, besides the fact
 that  $\Phi^{su}_{j;\pm j,\pm j } \Phi^{sl}_{j;-j-1,-j-1} \psi^\pm \bar \psi^\pm$
 do not preserve the same set of supercharges as  $ \Phi^{sl}_{0;- 1,-1}\psi^3 \bar \psi^3$ ,
 we cannot use them in any case to construct a real deformation that preserves
 supersymmetry. 
 
 \no
 The reason is that we cannot construct a real operator  by using
only the currents $J^+$ and $\bar J^+$ and, on the other hand,
operators with $\psi^+$ and its complex conjugate
 $\psi^-$ preserve complementary sets of supercharges, i.e.~($Q_{3\pm},Q_{4\pm}$)
and ($Q_{1\pm},Q_{2\pm}$) respectively, as we have already seen.
Therefore, 
we cannot construct a real supersymmetric deformation using these operators.
Two observations are now in order. First, this problem
does not arise when we use the $SL(2,\mathbb{R})$ fermions $\chi^+, \bar \chi^+$
because the currents $K^+, \bar K^+$ combine by themselves to
a real operator. Second, this issue did  not also arise in the setup of \cite{Prezas:2008ua}
because the analogues of
$\Phi^{su}_{j;\pm j,\pm j } \Phi^{sl}_{j;-j-1,-j-1} \psi^\pm \bar \psi^\pm$
preserve the full amount of supersymmetry (16 supercharges)
of the original undeformed NS5-brane configuration.

\no We would like to close this section with a final remark related to the fact that
the levels of both WZW models are identified with the number of NS5-branes $N_5$. As a consequence, the number of operators in the first expression in (\ref{opbr}) -- i.e. those who survive the reality condition and truly generate supersymmetric deformations -- scales approximately as $N_5^3$. From the  brane point of view we would have expected  $4 (N_1 + N_5)$ possible deformations, since we can move all branes arbitrarily. It is  not clear to us how this discrepancy should be interpreted (and eventually fixed), since the weak-string-coupling regime that guarantees the validity of the CFT analysis demands $N_1 \gg N_5$ (see (\ref{gstr})), which sets no order between  $N_5^3$ and  $4 (N_1 + N_5)$.

\section{More operators in the $SL(2,\mathbb{R}) \times SU(2) \times \mathbb{T}^4$ theory}

A large class of operators that give rise to marginal deformations
consists of bosonic primaries of the above models with vanishing
total conformal weight along with a fermion, so that the overall conformal weight is
$h=1/2$.
The simplest and most natural
construction involves two primaries. Therefore we have the following
two classes of operators. Either we use \eqn{susly}, that is
\begin{equation}
\label{firstclass}
{\rm First\ class}: \qq \Phi^{sl}_{j;m} \Phi^{su}_{j;n} {\cal Y}
\end{equation}
or
\begin{equation}
{\rm Second\ class}:\qq \Phi^{sl}_{j;m} e^{i p_a Y^a} {\cal Y}\ ,
\end{equation}
where ${\cal Y}$ denotes a fermion in one of the WZW models
or in the 4-torus and $Y^a, a=1,2,3,4$ are free bosons describing  the 4-torus. We will
also use the complex combinations $\hat Y^{\pm}= Y^1 \pm i Y^2$ and
$\tilde Y^{\pm}= Y^3 \pm i Y^4$ in the construction of the ${\cal N}=2$
superconformal algebra in appendix C.
The condition
$\frac{j(j+1)}{k} - \frac{1}{2} \sum_a p_a^2 =0$ should hold
(we consider for simplicity only momentum modes on the 4-torus) so that
the second class of operators have conformal weight $h=1/2$.
Notice that the momenta $p_a$ are quantized since the coordinates $Y^a$ are compact
but $j$ is an arbitrary real number in the range
\begin{equation}
-\frac{1}{2} \leqslant j \leqslant \frac{k-1}{2}\ .
\end{equation}

\subsection{First class}

Operators in the first class with the fermion ${\cal Y}$ being either
in the $SL(2,\mathbb{R})$ or the $SU(2)$ part of the theory
were studied in the previous section and we saw that several of them
can lead to deformations that preserve one-half of the original spacetime supersymmetry. For
this to happen the charge $n$ or $m$ from the WZW model, where ${\cal Y}$ belongs,
has to be fixed appropriately with respect to $j$. Furthermore,
some of these operators correspond to geometric deformations
of the F1-NS5-brane system similar to those studied in \cite{Fotopoulos:2007rm}.

\no
The operators (\ref{firstclass}) involving fermions from the 4-torus give rise
to deformations of the moduli of the torus which, in general, will depend on
the $SL(2,\mathbb{R})$ and $SU(2)$ coordinates through the
corresponding affine primaries.
We still have to check if any of those
give rise to supersymmetric deformations but since the form of these
deformations is not consistent with the general ansatz (\ref{generalmetric}),
we expect that none of those can preserve any supersymmetry.
It is an interesting exercise to see how this happens.

\no Let us start from the seed
operator
\begin{equation}
\label{fct}
\Phi^{sl}_{j;m} \Phi^{su}_{j;n} \hat \lambda^+\ ,
\end{equation}
where  we consider a specific complex fermion from the 4-torus
(obviously the analysis is similar for all
other 4-torus fermions). The deforming operator then reads
\begin{equation}
\Phi^{sl}_{j;m} \Phi^{su}_{j;n} \partial \hat Y^+\ ,
\end{equation}

\no
The associated fermion
bilinears and the supercharges that commute with them are
\begin{equation}
\label{}
\begin{array}{rcl}
\displaystyle{-(1+j-m) \Phi^{sl}_{j;m-1} \Phi^{su}_{j;n} \chi^+ \hat \lambda^+}&, \quad&\displaystyle{Q_{1+}\ ,\ \  Q_{2\pm}\ ,\ \
Q_{3\pm}\ ,\ \  Q_{4+}} \ , \crbig
\displaystyle{(j+n) \Phi^{sl}_{j;m} \Phi^{su}_{j;n-1} \psi^ + \hat \lambda^+}&, \quad&\displaystyle{Q_{1+}\ ,\ \ Q_{2+}\ ,\ \
Q_{3\pm}\ ,\ \  Q_{4\pm}} \ , \crbig
\displaystyle{-\sqrt{2} m \Phi^{sl}_{j;m} \Phi^{su}_{j;n} \chi^3 \hat \lambda^+}&, \quad&\displaystyle{ Q_{1+}\ , \ \  Q_{2+}\ ,\ \
Q_{3+}\ ,\ \  Q_{4+}} \ , \crbig
\displaystyle{ \sqrt{2} n \Phi^{sl}_{j;m} \Phi^{su}_{j;n} \psi^3 \hat \lambda^+}&, \quad&\displaystyle{Q_{1+}\ ,\ \ Q_{2+} \ ,\ \
Q_{3+}\ ,\ \  Q_{4+}} \ , \crbig
\displaystyle{(j-n) \Phi^{sl}_{j;m} \Phi^{su}_{j;n+1} \psi^- \hat \lambda^+}&, \quad&\displaystyle{ Q_{1\pm}\ ,\ \  Q_{2\pm}\ ,\ \
Q_{3+}\ ,\ \  Q_{4+}} \ , \crbig
\displaystyle{(1+j+m) \Phi^{sl}_{j;m+1} \Phi^{su}_{j;n} \chi^- \hat \lambda^+}&, \quad&\displaystyle{Q_{1\pm}\ ,\ \  Q_{2+}\ ,\ \
Q_{3+}\ ,\ \  Q_{4\pm}} \ .
\end{array}
\end{equation}
We notice that there is a common set of 4 commuting supercharges $Q_{i +}, i=1,\ldots, 4$
and that
therefore these operators preserve 1/2 of the original supersymmetry.
However, in order to get a real deformation we should add the complex
conjugates of the above deforming operators, which, as it can be easily seen,
 preserve the complementary set
$Q_{i -},i=1,\ldots,4$. Therefore, 
there are  no (real) supersymmetric deformations of this type, in accord with the fact that they are not expected
from the supergravity analysis. It is also elementary to show that the usual
moduli deformation of the torus, i.e.~of the form $\partial \hat Y^+  \bar \partial \hat Y^-
\pm \partial \hat Y^- \bar \partial \hat Y^+$, commute with all supercharges as they should.

\subsection{Second class}

The second class of operators, which contain a primary from the 4-torus,
leads also to 1/2 BPS deformations when ${\cal Y}$ is a fermion from
the $SL(2,\mathbb{R})$ model.
In particular, the following operators yield
deformations preserving 8 supercharges (including the holomorphic and
antiholomorphic sectors): $\Phi^{sl}_{j;\mp j \mp 1} e^{i p_a Y^a} \chi^{\pm}$.
The corresponding deformations contain the null currents of the $SL(2,\mathbb{R})$
WZW model and hence they reflect a situation where the harmonic
function $H_1$ in (\ref{generalmetric}) depends on the coordinates $y^a$ of the
4-torus.

\no
Let us find out which supercharges can be  preserved. We select as seed
\begin{equation}
\label{seedclb}
\Phi^{sl}_{j;m} e^{i p_a Y^a} \chi^{+}
\end{equation}
and we obtain the following 2-fermion terms in the associated deformation
\ba
&&a_b \Phi^{sl}_{j;m} e^{i p_a Y^a} \lambda^b \chi^+\ ,\nonumber
\\
&& \sqrt{2} (1+m)  \Phi^{sl}_{j;m} e^{i p_a Y^a} \chi^+ \chi^3 \ ,
\\
&& -(1+j+m)   \Phi^{sl}_{j;m} e^{i p_a Y^a} \chi^+ \chi^-\ .
\nonumber
\ea
Since no supercharges commute with the third term we have to set $m=-1-j$.
The second term preserves $Q_{2\pm}, Q_{3\pm}$ and these are also preserved
by the first term as well, due to the fact that they have $\epsilon_2=1$.
Therefore we have a 1/2 BPS deformation. Similarly, such operators
with $\chi^-$ preserve the complementary set $Q_{1\pm}$ and $Q_{4\pm}$ .

\no
Brane configurations  corresponding to deformations of the
F1-NS5-brane system driven by this type of operators
were studied in \cite{Tseytlin:1997cs}, where it was shown that
they are solutions of the equations of motion and preserve 1/4 of the original supersymmetry provided that
the harmonic condition on $H_1$ changes to
\begin{equation}
\big(\partial_x^2 + H_5(x) \partial_y^2 \big) H_1(x,y) =0\ .\label{tseytlinequ}
\end{equation}
A simple class of solutions of that equation, with $H_5$ being the standard
near-horizon form of the harmonic function on the transverse space
$H_5 = 1/r^2$, can be found by assuming a factorized form
of $H_1(x,y) = f(x) g(y)$. We get two equations
\begin{equation}
r^2 \frac{\partial_x^2 f(x)}{f(x)} = - \frac{\partial_y^2 g(y)}{g(y)} = c\ .
\end{equation}

\no
Since the coordinates $y^b$ parametrize 4-torus, the solution of the
second equation are of the form $g(y) = e^{i a_b y^b}$ with the condition
$c=\sum_b a_b^2$. Assuming furthermore that $f(x)$ depends only on the radial
coordinate $r$ yields
\begin{equation}
 r^2 f''(r) + 3 r f'(r) - c f(r) =0\ ,
 \end{equation}
with solutions $f(r)= r^{-1\pm \sqrt{1+c}}$. This solution is a deformation
of the original harmonic solution $H_1 =  1/r^2$.  Recall that the latter corresponds to
the F1-branes fully smeared on the 4-torus. The deformation reflects
a situation where
some momentum modes on the 4-torus are condensed and have to be compensated
by a change of the profile of the F1-branes. This change of profile can be thought
of as a deformation of the original smooth instanton to which the smeared
F1-branes correspond to. Therefore, these deformations trigger
infinitesimal motions in the instanton moduli space, the latter being
the Higgs branch of the F1-NS5-system.

\no
The conformal field theory description of these deformations is provided
by the operators $\Phi^{sl}_{j;-j-1} e^{i p_a Y^a} \chi^{+}$
which preserve the supercharges  $Q_{2\pm}$ and $Q_{3\pm}$.
These are exactly the supercharges preserved by the operators corresponding to
the circular and elliptical deformation studied previously,
in perfect agreement with the fact that the form of the Killing spinors does not
depend on the explicit form of the functions $H_1$ and $H_5$, even if we use the
more general ansatz of  \cite{Tseytlin:1997cs}.
The relation between the $a_b$ and $c$ is the classical analogue
of the quantum relation between $p_b$ and $j$ that results from
the condition of conformal invariance.
We should mention that operators of this type have not been considered so far
in discussions of the ${\rm \mathrm{AdS}}_3/ {\rm CFT}_2$ duality and it would be
very interesting to elucidate their role in that context.

\no
On the other side, it is not meaningful to give a $y$-dependence on the $H_5(x)$
harmonic function since it would imply a dependence of the harmonic function
describing the NS5-branes  on some of their worldvolume coordinates.
Therefore,  operators of the form (\ref{seedclb}) but with a fermion in the $SU(2)$
WZW model should not yield exactly marginal deformations. Here we will
restrict ourselves to showing that they cannot yield a real deformation that
preserves supersymmetry.

\no
Taking as seed the operator
\begin{equation}
 \Phi^{sl}_{j;m } e^{i p_a Y^a} \psi^+\ ,
 \end{equation}
yields the following fermion bilinears
\begin{equation}
\begin{array}{rcl}
&&\displaystyle{ a_b \Phi^{sl}_{j;m} e^{i p_a Y^a} \lambda^b \psi^+} \ , \crbig
&&\displaystyle{(1+j-m) \Phi^{sl}_{j;m-1} e^{i p_a Y^a}  \psi^+ \chi^+} \ , \crbig
&&\displaystyle{\sqrt{2} (m-1)  \Phi^{sl}_{j;m} e^{i p_a Y^a}  \psi^+ \chi^3} \ , \crbig
&&\displaystyle{ -\sqrt{2} (1+j+m)  \Phi^{sl}_{j;m+1} e^{i p_a Y^a}  \psi^+ \chi^- } \ .
\end{array}
\end{equation}
All these terms preserve simultaneously the supercharges $Q_{3\pm}$ and
$Q_{4\pm}$. However, as was the case with first class operators containing
$SU(2)$ fermions, 
in order to construct a real deformation we should also
add the complex conjugate operator that involves the fermion $\psi^-$
and these preserve the complementary set of supercharges 
$(Q_{1\pm}, Q_{2\pm})$, as it can easily be seen.

\no
Finally, operators of the second class with a fermion from the 4-torus are excluded
due to the same reason we excluded operators of the type (\ref{fct}).

\section*{Acknowledgements}

The authors wish to thank C.~Bachas, V.~Niarchos, and J.~Troost for stimulating discussions, and M.~Gaberdiel, I.~Kirsch, D.~Kutasov and  A.~Pakman for useful correspondence. They also acknowledge
A.~Fotopoulos for initial collaboration and helpful exchanges.
Marios Petropoulos thanks Neuch\^atel
University, Patras University, the  CERN Theory Division, the University
of Bern  and the Galileo Galilei Institute for Theoretical Physics for kind
hospitality at various stages of this collaboration, and
acknowledges financial support from the Swiss National Science
Foundation, the French Agence Nationale pour la Recherche, contract  05-BLAN-0079-01, and the
EU under the contracts  MRTN-CT-2004-005104
and MRTN-CT-2004-503369. The work of Nikolaos Prezas was
partially supported from the Swiss National Science
Foundation and he wishes to thank Patras University and the Ecole
Polytechnique for their warm hospitality. Kostandinos Sfetsos acknowledges the CPHT of the Ecole
Polytechnique and the LPT of the Ecole Normale Sup\'erieure for kind hospitality and  financial support from the
Groupement d'Int\'er\^et Scientifique P2I.

\appendix

\section{$SU(2)$ conventions}

\no
We use  the parametrization
of the $SU(2)$  matrix element employed in \cite{Fotopoulos:2007rm},
\begin{equation}
  \begin{pmatrix}
    \tilde g_{++}  & \tilde g_{+-} \\
    \tilde g_{-+} & \tilde g_{--}
  \end{pmatrix}
=
  \begin{pmatrix}
    \cos\th\ e^{i \phi } & \sin\th\ e^{i \tau } \\
    -\sin\th\ e^{- i \tau  } & \cos\th\
e^{-i \phi }
  \end{pmatrix} \ ,
\end{equation}
to obtain the following semiclassical expression for the primaries
 \be
 \Phi^{su}_{j;j,j}=\tilde g_{++}^{2j}\ ,\qq \Phi^{su}_{j;-j,-j}=\tilde
 g_{--}^{2j}\ , \qq \Phi^{su}_{j;j,-j}=\tilde g_{+-}^{2j}\ ,\qq
 \Phi^{su}_{j;-j,j}=\tilde g_{-+}^{2j}\ .
  \ee
The $SU(2)$ primary
$\Phi^{su}_{j;m,\bar m}$ at level $k-2$ has conformal weight
\begin{equation}
\Delta =  \frac{j (j+1)}{k}\ ,
\end{equation}
 with $j$ and $m$ being half-integers in the ranges
 \begin{equation}
0 \leqslant j \leqslant \frac{k-2}{2} \ , \quad
-j \leqslant m \leqslant j \ .
\end{equation}

\no
The left- and right-moving currents of the theory are given by 
\begin{eqnarray}
J^1 &=& 2 \big( \sin(\phi+\tau) \partial \theta+\cos (\phi+\tau)
\sin\theta \cos \theta (\partial \tau-\partial \phi)\big)\ , \nonumber\\
 J^2 &=& 2 \big(\cos(\phi+\tau) \partial \theta-\sin (\phi+\tau)
\sin\theta \cos \theta (\partial \tau-\partial \phi)\big)\ , \\
J^3 &=& 2 \big( \cos^2\theta \partial \phi+\sin^2 \theta \partial\tau\big)\ ,\nonumber
\end{eqnarray}
 and
 \begin{eqnarray}
\bar J^1 &=& -2 \big( \sin(\phi-\tau) \bar \partial \theta+\cos (\phi-\tau)
\sin\theta \cos \theta (\bar \partial \tau+\bar \partial \phi)\big)\ , \nonumber\\
 \bar J^2 &=& 2 \big(\cos(\phi-\tau) \bar \partial \theta+\sin (\phi-\tau)
\sin\theta \cos \theta (\bar \partial \tau+\bar \partial \phi)\big)\ , \\
\bar J^3 &=& 2 \big( \cos^2\theta \bar \partial \phi-\sin^2 \theta \bar \partial \tau\big)\ .\nonumber
\end{eqnarray}

\no
The action of the $SU(2)$ affine currents $J^3,
J^\pm=J^1 \pm i J^2$ on a primary field $\Phi^{su}_{j;m, \bar m}$
is given by the OPEs
\begin{equation}
\label{}
\begin{array}{rcl}
\displaystyle{J^3(z) \Phi^{{ su}}_{j;m, \bar m}(w, \bar w)}&=&\displaystyle{ \frac{m}{z-w}
\Phi^{su}_{j;m, \bar m}} (w, \bar w ) \
, \crbig \displaystyle{J^\pm(z) \Phi^{{ su}}_{j;m, \bar m}(w, \bar w)}&=&\displaystyle{ \frac{j\mp m}{z-w}
\Phi^{{su}}_{j;m\pm1, \bar m}} (w, \bar w) \ .
\end{array}
\end{equation}

\no
The bosonic current algebra reads
\begin{equation}
\label{SUalg}
\begin{array}{rcl}
\displaystyle{J^3(z)
J^3(w)}&\sim&\displaystyle{\frac{k-2}{2}\frac{1}{(z-w)^2}}\ , \crbig
\displaystyle{J^3(z)J^\pm(w)}&\sim&\displaystyle{\pm
\frac{J^\pm(w)}{z-w}}\ , \crbig
\displaystyle{J^+(z)J^-(w)}&\sim&\displaystyle{\frac{k}{(z-w)^2}+
\frac{2 J^3(w)}{z-w}}\ ,
\end{array}
\end{equation}
at level $k-2$. The corresponding
fermions satisfy
\begin{equation}
\label{SUfalg}
\begin{array}{rcl}
\displaystyle{\psi^3(z)
\psi^3(w)}&\sim&\displaystyle{\frac{1}{z-w}}\ , \crbig
\displaystyle{\psi^+(z)
\psi^-(w)}&\sim&\displaystyle{\frac{1}{z-w}}\ .
\end{array}
\end{equation}

\section{$SL(2,\mathbb{R})$ conventions}

 A  two-dimensional matrix realization of the $SL(2,\mathbb{R})$ algebra
is given in terms of Pauli matrices as follows:
\begin{align}
 k^1 =  \frac{i}{2} \sigma^2\ ,
   &&
 k^2 = - \frac{i}{2} \sigma^1\ ,
  &&
 k^3 =  \frac{1}{2} \sigma^3\ .
\end{align}
These matrices satisfy the $SL(2,\mathbb{R})$ commutation relations
\begin{align}
  \left[k^1 , k^2\right] = - i k^3\ , && \left[ k^2 , k^3\right] = i k^1 \ ,&&
  \left[ k^3 , k^1 \right] = i k^2\ ,
  \label{eq:commSL2}
\end{align}
and therefore
$k^\pm=k^1\pm i k^2$ and $k^3$ satisfy
\begin{align}
  \left[k^3 , k^+\right] =   k^+ \ , && \left[ k^3 , k^-\right] = -  k^- \ , &&
  \left[ k^+ , k^- \right] =- 2   k^3\ ,
\end{align}
which is the form of the $SL(2,\mathbb{R})$ we employ in the construction
of the ${\cal N}=2$ superconformal algebra.

\no
We will parametrize the matrix element of $SL(2,\mathbb{R})$ as
\begin{equation}
  g  =
  \begin{pmatrix}
    e^\rho   & e^\rho x^+ \\
    e^\rho x^- & e^{-\rho}+x^+ x^- e^\rho
  \end{pmatrix} =
  \begin{pmatrix}
    g_{++}  &  g_{+-} \\
    g_{-+} &  g_{--}
  \end{pmatrix}\ .
\end{equation}
The right-invariant 1-forms are
$j^a_{\mathrm{R}}=-i {\rm tr} (dg g^{-1} k^a)$ and they read
 \begin{eqnarray}
j^1_{\mathrm{R}} &=&  -\frac{i}{2} \Big( dx^-+ 2 x^- dr - e^{2\rho}\big(1+(x^-)^2\big) dx^+\Big)
\
,\nonumber \\
j^2_{\mathrm{R}} &=&- \frac{1}{2}\Big(dx^-  + 2 x^- dr + e^{2\rho} \big(1-(x^-)^2\big) dx^+\Big)
 \ , \\
j^3_{\mathrm{R}} &=&  i (e^{2 \rho} x^- dx^+ - dr) \ ,\nonumber
\end{eqnarray}
while the left-invariant 1-forms $j^a_{\mathrm{L}} = - i  {\rm tr} (g^{-1} dg  k^a)$
read
\begin{eqnarray}
 j^1_{\mathrm{L}} &=&   \frac{i}{2}\Big(dx^+ +2 x^+ dr -e^{2 \rho}\big(1+(x^+)^2\big)d x^-\Big)
\
,\nonumber \\
 j^2_{\mathrm{L}} &=&  -\frac{1}{2}( dx^+ + 2 x^+ dr + e^{2\rho} \big(1-(x^+)^2\big) dx^- \Big)
  , \\ j^3_{\mathrm{L}} &=& i (e^{2\rho} x^+ dx^- - dr) \ .\nonumber
\end{eqnarray}
The Cartan--Killing metric $ds^2_{\mathrm{CK}}=(j^1_{\mathrm{R}})^2+(j^2_{\mathrm{R}})^2-(j^3_{\mathrm{R}})^2=(j^1_{\mathrm{L}})^2+(j^2_{\mathrm{L}})^2-(j^3_{\mathrm{L}})^2$ is
\begin{equation}
ds^2 = d\rho^2 + e^{2\rho} dx^+ dx^-  \ .
\end{equation}
  Consequently, the
 left- and right-moving currents of the WZW model are
 \begin{eqnarray}
K^1 &=& -\frac{i}{2} \Big( \partial x^-+ 2  x^- \partial \rho - e^{2\rho}\big(1+(x^-)^2\big) \partial x^+\Big)  \
,\nonumber \\
K^2 &=&   - \frac{1}{2}\Big(\partial x^-  + 2 x^- \partial \rho + e^{2\rho} \big(1-(x^-)^2\big) \partial x^+\Big)
 \ , \\
K^3 &=&  i (e^{2 \rho} x^- \partial x^+ - \partial \rho)   \ ,\nonumber
\end{eqnarray}
and
\begin{eqnarray}
\bar K^1 &=&  \frac{i}{2}\Big(\bar \partial x^+ +2 x^+ \bar \partial \rho -e^{2 \rho}\big(1+(x^+)^2\big)\bar \partial x^-\Big)
\
,\nonumber \\
\bar K^2 &=&  -\frac{1}{2}( \bar \partial x^+ + 2 x^+ \bar \partial \rho + e^{2\rho} \big(1-(x^+)^2\big)\bar \partial x^- \Big)
 , \\
\bar K^3 &=& i (e^{2\rho} x^+ \bar  \partial x^- - \bar \partial \rho)
 \ ,\nonumber
\end{eqnarray}
respectively.

 \no
Now we can identify the charges of the combinations $K^1\pm i K^2$
by using the Killing vector fields $ j_{i{\mathrm{R}}},$ and $ j_{i{\mathrm{L}}}$
dual to the forms
$ j^i_{{\mathrm{R}}},$ and $ j^i_{{\mathrm{L}}}$. It turns out that

\begin{equation}
\begin{array}{rcl}
\displaystyle{ [ j_{3{\mathrm{R}}}, j_{1{\mathrm{R}}} \pm i j_{2{\mathrm{R}}} ] }&=&\displaystyle{\mp 2 i ( j_{1{\mathrm{R}}} \pm i j_{2{\mathrm{R}}})} \ , \crbig
\displaystyle{ [j_{3{\mathrm{L}}}, j_{1{\mathrm{L}}} \pm i j_{2{\mathrm{L}}} ]}&=&\displaystyle{ \pm 2 i ( j_{1{\mathrm{L}}} \pm i j_{2{\mathrm{L}}})} \ .
\end{array}
\end{equation}
Therefore, to be consistent with the way we picked up the charges
of the $SL(2,\mathbb{R})$ currents in our construction of the
${\cal N}=2$ SCFT algebra in appendix C, we should define $K^\pm=K^1\mp i K^2$
and $\bar K^\pm = \bar K^1 \pm i \bar K^2$.  We have
\begin{equation}
K^+ = i e^{2\rho} \partial x^+, \quad K^- = -i \Big(\partial x^- + 2 x^- \partial \rho
- e^{2\rho} (x^-)^2 \partial x^+\Big)\ ,
\end{equation}
and
\begin{equation}
\bar K^+ = - i e^{2\rho} \bar \partial x^-, \quad \bar K^- = i \Big (\bar \partial x^+ + 2
x^+ \bar \partial \rho - e^{2\rho} (x^+)^2 \bar \partial x^-\Big)\ .
\end{equation}
Hence one obtains the following useful relation
\begin{equation}
\partial x^+ \bar \partial x^- =
\Phi^{sl}_{1;-2,-2} K^+ \bar K^+ \ ,
\end{equation}
where we used the semiclassical expressions for the $SL(2,\mathbb{R})$ primaries
\begin{equation}
\begin{array}{rclrcl}
\Phi^{sl}_{j;j+1,j+1}&=&\displaystyle{{1\ov g_{--}^{2(j+1)}}}\   ,
&\quad \Phi^{sl}_{j;-j-1,-j-1}&=&\displaystyle{{1\ov g_{++}^{2(j+1)}}}\
, \crbig \Phi^{sl}_{j;j+1,-j-1}
 &=&\displaystyle{{1\ov g_{-+}^{2(j+1)}}}\ , &\quad
 \Phi^{sl}_{j;-j-1,j+1}&=&\displaystyle{{1\ov g_{+-}^{2(j+1)}}}\ .
\end{array}
\end{equation}

 \no
Notice that we use conventions where the $SL(2,\mathbb{R})$ primary
$\Phi^{sl}_{j;m,\bar m}$ at level $k+2$ has conformal weight  \begin{equation}
\Delta = - \frac{j (j+1)}{k}\ .
\end{equation}
We consider only the principal discrete series for which
 $j$ is a real number, since we actually consider the universal cover
 of $SL(2,\mathbb{R})$ in order to avoid any closed timelike curves,
in the range
 \begin{equation}
-\frac{1}{2} \leqslant j \leqslant \frac{k-1}{2} \ ,
\end{equation}
and $m$ takes either the values  $m=-j-j,-j-2,\ldots$ or
$m=j+1,j+2,\ldots$.

\no
The action of the $SL(2,\mathbb{R})$ affine currents $K^3,K^\pm$
on a primary field $\Phi^{sl}_{j;m, \bar m}$
is given by the OPEs
\begin{equation}
\label{}
\begin{array}{rcl}
\displaystyle{K^3(z) \Phi^{{ sl}}_{j;m, \bar m
}(w,\bar w )}&=&\displaystyle{ \frac{m}{z-w} \Phi^{sl}_{j;m, \bar m}} (w,\bar w) \
, \crbig \displaystyle{K^\pm(z) \Phi^{{ sl}}_{j;m, \bar
m}(w, \bar w)}&=&\displaystyle{ \frac{m\pm(j+1)}{z-w}
\Phi^{{sl}}_{j;m\pm1, \bar m}} (w, \bar w) \ .
\end{array}
\end{equation}

\no
The bosonic current  algebra at level $k+2$ reads
\begin{equation}
\label{SLalg}
\begin{array}{rcl}
\displaystyle{K^3(z)
K^3(w)}&\sim&\displaystyle{-\frac{k+2}{2}\frac{1}{(z-w)^2}}\ ,
\crbig \displaystyle{K^3(z)K^\pm(w)}&\sim&\displaystyle{\pm
\frac{K^\pm(w)}{(z-w)}}\ , \crbig
\displaystyle{K^+(z)K^-(w)}&\sim&\displaystyle{\frac{k+2}{(z-w)^2}-2
\frac{ K^3(w)}{z-w}}\ ,
\end{array}
\end{equation}
while the
corresponding fermions satisfy
\begin{equation}
\label{SLfalg}
\begin{array}{rcl}
\displaystyle{\chi^3(z)
\chi^3(w)}&\sim&-\displaystyle{\frac{1}{z-w}}\ , \crbig
\displaystyle{\chi^+(z)
\chi^-(w)}&\sim&\displaystyle{\frac{1}{z-w}}\ .
\end{array}
\end{equation}

\section{The $\mathcal{N}=2$ superconformal algebra}

\textbf{We present here a realization of the $\mathcal{N}=2$ superconformal algebra in the $SL(2,\mathbb{R})\times SU(2)\times U(1)^4$ worldsheet theory.}
 The energy--momentum tensor reads
\begin{eqnarray}\label{emt}
  T&=&\frac{1}{k}\left[
J^3 J^3 +\frac{1}{2}\left( J^+ J^- + J^-J^+ \right)- K^3 K^3
+\frac{1}{2}\left( K^+ K^- + K^-K^+ \right)
  \right]+\frac{1}{2} \sum_{a=1}^4 \partial Y^a \partial Y^a
  \nonumber\\
&&-\frac{1}{2}\left[\psi^+\partial\psi^- +\psi^-\partial\psi^+
+\psi^3\partial\psi^3+\chi^+\partial\chi^-
+\chi^-\partial\chi^+ -\chi^3\partial\chi^3+  \sum_{a=1}^4 \lambda^a \partial \lambda^a
\right]\ ,
\end{eqnarray}
while the ${\cal N}=2$ supercurrents take the form
\begin{equation}
\label{emtpart}
\begin{array}{rcl}
\displaystyle{ G^+}&=&\displaystyle{ \frac{1}{\sqrt{k}}\left[\left( J^3_{\mathrm{T}}+ K^3_{\mathrm{T}}\right)
\left(\psi^3-\chi^3\right) + \sqrt{2}\left(J^+ \psi^-+ K^-
\chi^+\right) \right]+ \hat \lambda^+ \partial \hat Y^- + \tilde \lambda^+ \partial \tilde Y^-} \ , \crbig
\displaystyle{G^-}&=&\displaystyle{ \frac{1}{\sqrt{k}}\left[\left( J^3_{\mathrm{T}}- K^3_{\mathrm{T}}\right)
\left(\psi^3+\chi^3\right) + \sqrt{2}\left(J^- \psi^++ K^+
\chi^-\right)\right]
+ \hat \lambda^- \partial \hat Y^+ + \tilde \lambda^- \partial \tilde Y^+} \ .
\end{array}
\end{equation}
The $U(1)$ $R$-charge current reads
\begin{equation}
\label{emtcur}
  J_R= \frac{2}{k}\left(J^3_{\mathrm{T}}+K^3_{\mathrm{T}}\right)
 - \psi^+\psi^- + \chi^+\chi^-  + \psi^3\chi^3 + \hat \lambda^+  \hat \lambda^- +
  \tilde \lambda^+  \tilde \lambda^-\ ,
\end{equation}
where we have introduced the total $SU(2)$ and $SL(2,\mathbb{R})$
currents
\begin{equation}\label{totcur}
  J^3_{\mathrm{T}}= J^3 + \psi^+\psi^-\ , \qq
  K^3_{\mathrm{T}}= K^3 + \chi^+\chi^-\ .
\end{equation}

\textbf{ Due to the presence of a time-like direction in the interacting non-linear sigma-model, the above generators turn out to be non-hermitian: the usual complex conjugation between $G^+$ and $G^-$ does not hold.
}

\end{document}